\documentclass[12pt]{article}

\title{\bf
Scaling Analysis of the Site-Diluted Ising Model in Two Dimensions}
\author{ 
{\it R. Kenna}\\
Applied Mathematics Research Centre,\\
Coventry University,\\
Coventry, CV1 5FB, England
{}\\~\\and \\~\\
{\it J.J. Ruiz-Lorenzo}\\ 
Departamento de Fisica,\\
 Universidad de Extremadura,\\
Avda Elvas s/n, \\
Badajoz, 06071 Spain. }
\begin{document}
\maketitle
                      {\Large
                      \begin{abstract}
%
A combination of recent numerical and theoretical advances are applied to analyze the scaling behaviour of the
site-diluted Ising model in two dimensions, paying special attention to the implications for multiplicative
logarithmic corrections. The analysis focuses primarily on the odd sector of the model (i.e., that associated with magnetic exponents), 
and in particular on its Lee-Yang zeros, which are determined to high accuracy. 
Scaling relations are used to connect to the even (thermal) sector, and a first analysis of the 
density of zeros yields information on the specific heat and its corrections. 
The analysis is fully supportive of the strong scaling hypothesis and of the
scaling relations for logarithmic corrections.
%
                        \end{abstract} }
%
  \thispagestyle{empty}
%
%
  \newpage
%
                  \pagenumbering{arabic}

\section{Introduction}
\label{intro}
\setcounter{equation}{0}

The Ising model in two dimensions with uncorrelated, quenched random-site or random-bond disorder is a paradigm 
for the study of the statistical mechanics of disordered systems. 
With moderate dilution, the random system exhibits a phase transition different to that of the pure system, 
and the nature of this transition has been investigated for over three decades.

For the disordered Ising models in two dimensions, two scenarios have arisen. 
The strong universality hypothesis maintains that the leading critical exponents remain 
the same as in the pure case and that  the disorder induces multiplicative logarithmic corrections to scaling,
while the weak universality hypothesis favours dilution-dependent leading critical exponents.
While the former is now mostly favoured, especially in the random-bond Ising model (RBIM),
the debate regarding the scaling behaviour of the specific heat has persisted, especially in the site-diluted Ising model (RSIM),
focusing on whether this is characterised by an extremely weak double-logarithmic divergence or a finite
cusp. 
Indeed, according to the Harris criterion~\cite{Ha74}, the vanishing of the specific-heat  critical exponent $\alpha$ in the pure Ising model,
marks the borderline between the 
$\alpha > 0$ case, where disorder is relevant and where the critical exponents may change as random quenched disorder is added,
and the $\alpha < 0$ scenario, where this type of disorder does not alter critical behaviour and the critical exponents 
are unchanged~\cite{Ha74}. 
In this borderline circumstance, logarithmic corrections to scaling (distinct from the logarithmic divergence in the 
specific heat in the pure Ising system)  may arise. 

The issue of scaling in the RSIM is addressed here, as that is the more contentious version of the model.
All of the thermodynamic information concerning a statistical mechanical system is contained in the locus and density 
of its partition function zeros~\cite{LY}.
In   Ref.~\cite{RL}, the first numerical analysis of the Lee-Yang zeros in a disordered system (the RSIM) was performed 
 and 
the leading critical behaviour of the lowest lying Lee-Yang zeros was extracted
through a finite-size scaling (FSS) analysis.
A robust numerical technique to construct the density of zeros 
from simulational data was given in Ref.~\cite{density}. Finally, a self-consistent scaling theory which links the
exponents characterising the logarithmic corrections to scaling was recently presented in Refs.~\cite{KeJo06a,KeJo06b}.

Here, these numerical \cite{RL,density} and analytical \cite{KeJo06a,KeJo06b} advances are combined
to unambiguously determine the leading scaling behaviour and potential multiplicative logarithmic
corrections in the specific heat through the density of Lee-Yang zeros and the scaling relations.
This and all multiplicative logarithmic correction-to-scaling exponents for FSS of odd functions are zero.
I.e., there are no multiplicative logarithmic corrections for the FSS of the magnetic susceptibility, the individual  Lee-Yang zeros
and the density of zeros. This comes about through the delicate manner in which the exponents 
of the logarithms, which are nonzero in thermal scaling, balance each other out.
In this way,  it is established that the Lee-Yang zeros of disordered systems can be precisely determined numerically,
 a density-of-zeros analysis is applicable to such a system, also at the level of logarithms,  new scaling relations for the logarithmic
corrections \cite{KeJo06a,KeJo06b}  in this model are confirmed 
and a negative exponent for the contentious specific heat or its multiplicative 
logarithmic correction is made unlikely.

\section{Logarithmic Corrections and Scaling Scenarios}
\label{Logs}
\setcounter{equation}{0}

Recently a self-consistent scaling theory for logarithmic-correction exponents has been presented \cite{KeJo06a,KeJo06b}. 
Denoting the reduced temperature by $t$ and the reduced external field by $h$, this theory deals with the circumstances 
where, in the absence of field, the specific heat, magnetization, susceptibility and correlation length scale respectively as 
\begin{eqnarray}
  C_\infty (t) & \sim & |t|^{-\alpha} |\ln{|t|}|^{\hat{\alpha}}
\,,
\label{C}
\\
  m_\infty (t) & \sim & |t|^{\beta} |\ln{|t|}|^{\hat{\beta}}
\,,
\label{m}
\\
\chi_\infty (t) & \sim & |t|^{-\gamma} |\ln{|t|}|^{\hat{\gamma}}
\,,
\label{chi}
\\
\xi_\infty (t) & \sim & |t|^{-\nu} |\ln{|t|}|^{\hat{\nu}}
\,.
\label{xi}
\end{eqnarray}
At $t=0$ the magnetization scales with reduced field as
\begin{equation}
  m_\infty (h)  \sim  h^{\frac{1}{\delta}} |\ln{|t|}|^{\hat{\delta}}
\,.
\label{mh}
\end{equation}
The locus of Lee-Yang zeros along the imaginary $h$-axis is parameterized by $r$
and the  Yang-Lee edge, which marks the end of their distribution, is  denoted by $r_{\rm{YL}} (t)$.
In the symmetric phase, this scales as
\begin{equation}
  r_{\rm{YL}} (t)  \sim  |t|^{\Delta} |\ln{|t|}|^{\hat{\Delta}}
\,,
\label{r}
\end{equation}
while the density of these zeros along their locus ($r > r_{\rm{YL}}$) at criticality ($t=0$) behaves as \cite{density}
\begin{equation}
  g_{\infty} (r)  \sim  r^{a_2-1} |\ln{r}|^{\hat{a}_2}
\,.
\label{G}
\end{equation}
The leading critical exponents are related by the standard scaling relations for continuous phase transitions \cite{history}.
The theory presented in \cite{KeJo06a,KeJo06b} relates the exponents of the logarithmic corrections in an analogous manner.

With the strong universality hypothesis, the leading critical exponents for the dilute Ising models are identical to their pure 
counterparts: 
\begin{equation}
 \alpha = 0\,, \quad 
 \beta = \frac{1}{8}\,,\quad 
 \gamma = \frac{7}{4}\,,\quad 
 \delta = 15\,,\quad 
 \nu = 1\,,
\label{leading}
\end{equation}
with gap exponent 
\begin{equation}
 \Delta = \beta + \gamma =  \frac{15}{8}  \,.
\label{gap} 
\end{equation}
The latter equation has been verified in Ref.~\cite{RL}. 
The exponent $a_2$ characterizing the leading behaviour of the density 
of Lee-Yang zeros in Eq.(\ref{G}) is \cite{KeJo06a}
\begin{equation}
 a_2 
        = \frac{2 - \alpha}{\Delta}
\,,
\label{a2}
\end{equation}
which gives $a_2=16/15$ if the strong hypothesis (\ref{leading}) and (\ref{gap}) holds.

Shalaev \cite{Boris} and later  Shankar and Ludwig \cite{SL},  and then Jug and Shalaev \cite{JuSh96}, used 
field theory, bosonization techniques and conformal invariance                              
to derive theoretical predictions for the logarithmic-correction exponents in the 
random-bond case with sufficiently small quenched  dilution.
These corrections, which we term the SSLJ exponents, are  \cite{Boris,SL,JuSh96}
\begin{equation}
 \hat{\alpha} = 0\,, \quad 
 \hat{\beta} = -\frac{1}{16}\,,\quad 
 \hat{\gamma} = \frac{7}{8}\,,\quad 
 \hat{\delta} = 0\,,\quad 
 \hat{\nu} = \frac{1}{2}\,
\label{SSLJscaling}
\end{equation}
with specific heat diverging as a double logarithm \cite{KeJo06b,Boris,SL,DD}
\begin{equation}
  C_\infty (t) \sim  |\ln{|\ln{|t|}|}|
\,.
\label{loglog}
\end{equation}

The questions addressed in the literature over the past three decades  concerned
(i) the validity of the strong scaling hypothesis and the veracity of the leading exponents (\ref{leading}) in the
diluted cases, 
(ii) the validity of the theoretical derivation of the SSLJ correction exponents (\ref{SSLJscaling}), 
which relies on certain assumptions  regarding  the nature of the dilution
in the random-bond case (see, also, Ref.~\cite{Pl98} for 
the random-site case),
(iii) whether these exponents are dilution independent, 
(iv) the vanishing of $\hat{\alpha}$ and the validity of the double logarithm in (\ref{loglog}), and
(v) whether or not these
sets of exponents also apply to the random-site version of the model.

For the two-dimensional random Ising models, an alternative scenario to (i) has persisted in the literature. 
This is the weak universality hypothesis and maintains that certain leading critical exponents change 
continuously as the concentration of impurity defects is increased \cite{Su74}. 
In particular, the exponents $\alpha$, $\beta$, $\gamma$ and $\nu$ are maintained to be dilution dependent 
while $\delta$, $\eta$ and the ratios $\beta/\nu$ and  $\gamma / \nu$ remain independent of the dilution.
Furthermore, while agreeing with the SSLJ double-logarithmic  form (\ref{loglog})
for the specific heat,
Dotsenko and Dotsenko (DD) used the renormaliztion group  to predict~\cite{DD}
\begin{equation}
 \chi_\infty(t) \sim t^{-2} \exp{\left(-c(\ln{(-\ln{t})})^2\right)}
\,,
\label{chiDD}
\end{equation}
where $c$ is a constant related to the concentration of disorder.
While the SSLJ and DD predictions for the susceptibility differ substantially, 
it is fair to say that after much work by various authors
the  strong hypothesis is now mostly favoured (especially in the random-bond model).

The validity of the logarithmic-correction exponents (\ref{SSLJscaling}) proved harder to establish quantitatively. 
The first direct, clear, quantitative validation of the SSLJ prediction for the magnetic susceptibility (that 
$\hat{\gamma} = 7/8$) came in Ref.~\cite{RoAd98} through series expansions.
While the detailed scaling behaviour of  (\ref{chiDD})  has long been ruled out,  
the weak versus strong controversy persisted.
In contrast to the susceptibility, DD and SSLJ agree on the double logarithmic behaviour of the specific heat.
However, this has been notoriously difficult to confirm numerically, and has been the source of much controversy.

\begin{table}
\caption{Selection of recent  works supportive of  the weak or strong scaling hypothesis. 
}  
\begin {center}
\begin{tabular}{ll|l|l}  \hline \hline
                                        &                                                                      & RBIM                                                                             & RSIM \\
\hline
\multicolumn{2}{l|}{Support strong universality hypothesis }      &  \cite{RoAd98,BeCh04,Aade96,de97Lede06,SzIg99LaIg00}   & \cite{Se94,deSt94,ShVa01}  \\
 & $\quad$and theoretical support for $\alpha = \hat{\alpha}=0$ &  \cite{Boris,SL,JuSh96,DD}                                          &  \cite{Pl98} \\
 & $\quad$or numerical  support for $\alpha = \hat{\alpha}=0$     &   \cite {AnDo90WaSe90TaSh94,WiDo95,Aade97StAa97,HaTo08}&  \cite{HaTo08,BaFe97,SeSh98} \\
\hline
\multicolumn{2}{l|}{Support for weak universality hypothesis }              &   \cite{Ki95}                            & \cite{FaHo92} \\
 & $\quad$and theoretical support for finite $C_\infty(t)$                         &  \cite{finiteCinBond,Zi91}       &  \cite{Zi88} \\
 & $\quad$or numerical  support  for finite $C_\infty(t)$                          &  \cite{Ki00b,Ki00}                            &\cite{He92,KiPa94,Ku94KuMa00,HaMa08} \\
\hline \hline
\end{tabular}
\end{center}
\label{tab1}
\end{table}
Simulational works, generally supportive of the vanishing of $\alpha$ and $\hat{\alpha}$ and the 
specific heat diverging as a double logarithm viz. (\ref{loglog}),  
are found in Refs.~\cite{AnDo90WaSe90TaSh94,WiDo95,Aade97StAa97,HaTo08} 
and  Refs.~\cite{HaTo08,BaFe97,SeSh98} for the bond-disordered and random-site Ising models, respectively
(see also Refs.~\cite{Pl98,RoAd98,BeCh04,Aade96,de97Lede06,SzIg99LaIg00,Se94,deSt94,ShVa01}).
Indeed, plots of the measured specific heat as a function of the double logarithm of the lattice extent are contained in
Refs.~\cite{AnDo90WaSe90TaSh94,HaTo08} and Refs.~\cite{HaTo08,BaFe97,SeSh98} for the RBIM and the RSIM, respectively.
However, in  \cite{Ki00b} it was claimed that such apparent double-logarithmic FSS behaviour
does not necessarily imply divergence of the specific heat,
and numerically based counter-claims that the specific heat remains finite (so that $\alpha<0$ or 
$\alpha = 0$ and $\hat{\alpha}<0$)
in the random-bond \cite{Ki00b,Ki00}    and random-site models \cite{He92,KiPa94,Ku94KuMa00,HaMa08} also exist
(see also Refs.~ \cite{Ki95,FaHo92,finiteCinBond,Zi91,Zi88}).
The situation is summarized in Table~1.

The difficulties in unambiguously discriminating between the weak and strong scenarios on the basis of finite-size data were highlighted in
Refs.~\cite{Ki00b,Ku94KuMa00,MaKu99}. Indeed, in  Ref.~\cite{RoAd98} a fit to 
\begin{equation}
c_\infty(t) \propto |\ln{t}|^{\tilde{\alpha}}
\,,
\label{other}
\end{equation}
was attempted, and it was observed that
$\tilde{\alpha}$ decreases from the value of $1$ 
(which corresponds to the pure model) as the strength of disorder is increased.
This could be interpreted as supporting almost any reasonable value $\tilde{\alpha} \le 1$.
In Ref.~\cite{Ku94KuMa00} it was pointed out that specific heat data in the literature, which were stated to be supportive of
(\ref{loglog}), can often equally be fitted to (\ref{C}) with negative $\alpha$, still consistent with the Rushbrooke relation.

In this paper we address the problem from a fresh perspective, namely
that of partition function zeros (see also Ref.~\cite{RL}).  
In Ref.~\cite{KeJo06a} the scaling relation
\begin{equation}
 \hat{\Delta} = \hat{\beta} - \hat{\gamma}
\,,
\label{gaplog}
\end{equation}
for the logarithmic correction to the FSS of the Lee-Yang zeros was derived. 
For the diluted Ising models in two dimensions, this leads to the prediction
\begin{equation}
 \hat{\Delta} = -\frac{15}{16} = - 0.9375 
\,,
\label{loggap}
\end{equation}
and verification of this value for the logarithmic correction exponent characterising  the scaling of the Yang-Lee edge
in (\ref{r}) is one of the aims of this work.

In addition to 
testing the applicability and efficacy of the Lee-Yang-zero technique
in random models, also at the level of logarithmic corrections, a
central aim of this paper is to measure the density of zeros.
Indeed, the  correction exponent for the density of zeros is given in Ref.~\cite{KeJo06a} as
\begin{equation}
                         \hat{a}_2 = \frac{{\gamma \hat{\Delta}} + \Delta \hat{\gamma}}{\Delta} 
 \,.
\end{equation}
From the scaling relations for logarithmic corrections, 
in the special circumstances which prevail the diluted Ising models in two dimensions 
given in Ref.~\cite{KeJo06b},  $\hat{a}_2$ is 
related to the specific-heat correction exponent $\hat{\alpha}$ appearing in (\ref{C}) via 
\begin{equation}
 \hat{\alpha}  =   1 + 2\frac{\hat{\Delta}}{\Delta} + {\hat{a}}_2     =  1  +  \frac{16}{15}\hat{\Delta}+{\hat{a}}_2\, ,
\label{uncommon}
\end{equation}
having used the established value (\ref{gap}) for the leading gap exponent \cite{RL}. 
The elusive specific heat scaling exponents  $\alpha$, $\hat{\alpha}$ can thus be measured  
from the density  (\ref{G}) together with (\ref{a2}) and (\ref{uncommon}) and, in this way, one can distinguish between the competing $\alpha=0$, $\hat{\alpha}=0$ and
$\alpha<0$ or $\hat{\alpha}<0$ scenarios.

Since it is more contentious, we address
the site-diluted version.  Unlike the self-dual random-bond version, the
location of the critical temperature in the site-diluted Ising model  is not
exactly known  and has to be estimated numerically. In this work,
the highly accurate measurements for the critical temperatures reported in Ref.~\cite{BaFe97} for
different values of the site dilution are used.

\section{Simulation of the RSIM}
\label{Simulation}
\setcounter{equation}{0}

The partition function for a given realization of the RSIM in a reduced magnetic field $h$ is
\begin{equation}
 Z_L(\beta,h) = 
 \sum_{\{\sigma_i\}}{\exp{\left( \beta \sum_{\langle{i j  }\rangle}{\epsilon_i \epsilon_j\sigma_i \sigma_j} + h \sum_i{\epsilon_i\sigma_i}\right)}}
\,,
\label{pf}
\end{equation}
where $L$ denotes the linear extent of the lattice and the sum over configurations $\{\sigma_i\}$ is taken over Ising spins $\sigma_i \in \{\pm1\}$
and where $\epsilon_i$ are independent quenched random variables which take the  values unity with probability $p$ and zero with probability $1-p$.
For simulational purposes a regular (square) lattice with periodic boundary conditions is used.
The percolation threshold for such a lattice in the thermodynamic limit 
 occurs at $p=p_c = 0.592\,746\dots$, so that for $p<p_c$ the lattice fragments into finite-size
systems on which no true transition can occur \cite{percolation}.
Writing
\begin{equation}
 S= \sum_{\langle{i j  }\rangle}{\epsilon_i \epsilon_j\sigma_i \sigma_j}\,,
 \quad \quad \quad 
 M = \sum_i{\epsilon_i\sigma_i} 
\,,
\end{equation}
and 
\begin{equation}
 \rho_L(\beta;M) = \sum_S{\rho_L(S,M)\exp{\beta S}}
\,,
\end{equation}
where the spectral density $\rho_L(S,M)$ gives the relative weight of configurations with given values of $S$ and $M$,
the partition function in imaginary field $ih$ is 
\begin{equation}
 Z_L(\beta,h) = 
 \sum_M{\rho_L(\beta;M) \exp{(ihM)}}
 = 
  Z_L(\beta,0) 
 \left\langle{
   \cos{(hM)} + i \sin{(hM)}
 }\right\rangle
\,,
\label{pf2}
\end{equation}
where the expectation value has real measure. 
\begin{table}
\caption{The number of samples simulated for each lattice size $L$ at each 
value of the site-occupation probability and corresponding $\beta_c$ value (from Ref.~\cite{BaFe97}). }  
\begin {center}
\begin{tabular}{ll|l|l|r|r|r|r|r|r|}  \hline \hline
$L$ & & 32 & 48 & 64 & 96 & 128 & 196 & 256 \\
\hline
$p=0.88889$  & $\beta _c= 0.53781$ & 1000 & 1000 & 1000 & 600 & 600 & 250 & 250 \\
$p=0.75      $  & $\beta_c = 0.77125$ &1000 &1000 &1000 &1000 & 1000 &550 &270 \\
$p=0.66661$  & $\beta_c = 1.10$ & 1000 & 1000 & 1000 & 920 & 600 & 270
& 160  \\
\hline \hline
\end{tabular}
\end{center}
\label{tab2}
\end{table}
Assuming the Lee-Yang theorem holds \cite{LY,RL},
since odd moments of the magnetization vanish for $t\ge0$ ($\beta \le \beta_c$), the  zeros for a given realization of disorder
are given by the 
values of $h$ for which 
\begin{equation}
 \left\langle{
   \cos{hM} 
 }\right\rangle = 0
\,.
\end{equation}
Finally, for each value of $L$ and $p$, these zeros are averaged over realizations of disorder and the resulting $j^{\rm{th}}$
Lee-Yang zero is denoted by $r_j(L)$. Errors associated with the zeros are computed as sample-to-sample fluctuations.

We have simulated, using the Wolff single-cluster algorithm \cite{Wolff89}, three
different values of the dilution, namely $p=0.88889$, $0.75$ and $0.66661$.
From Ref.~\cite{BaFe97} the values of the critical temperatures for these three dilutions
are $\beta_c=0.53781(2)$ for $p=0.88889$, $\beta_c=0.77125(8)$ for $p=0.75$,
while $\beta_c=1.10$ corresponds to  $p=0.66661(3)$.  
In each of these three cases we have run
lattices of extent $L=32$, 48, 64, 96, 128, 196 and 256. 
The number of samples simulated for each lattice size and each value of the site-occupation probability
(and corresponding $\beta_c$ value) is given in Table~\ref{tab2}.

We have monitored the behavior of the non-local observables (such as the
susceptibility) with the Monte Carlo time by using a standard
logarithmic binning of the dynamical data. We have checked that, in
all the cases, the non-local observables have reached a plateau (as a
function of the Monte Carlo time).

\section{Scaling and Density Analyses}
\label{Simulation}
\setcounter{equation}{0}

The analysis focuses on the odd sector of the RSIM, which is connected to the even sector through the standard scaling relations
and their logarithmic counterparts \cite{KeJo06a,KeJo06b}. 
This connection is used to determine the specific-heat exponents through the density of zeros, 
in addition to a detailed analysis of the susceptibility and the Lee-Yang zeros.

The FSS analyses of the magnetic susceptibility and Yang-Lee edge
focuses on the correction-to-scaling exponents.
From (\ref{xi}), the reduced temperature is expresed in terms of the correlation length near criticality as
\begin{equation}
 t \sim \xi_\infty^{-\frac{1}{\nu}}\left({ \ln{\xi_\infty} }\right)^{\frac{\hat{\nu}}{\nu}}
\,.
\label{t}
\end{equation}
Substituting (\ref{t}) into (\ref{chi}) and (\ref{r}), gives the scaling behaviour for
susceptibility and the lowest lying Lee-Yang zeros in terms of the correlation length.
Because there are no logarithmic corrections to the FSS behaviour of the correlation length  \cite{KeJo06b}, 
for sufficiently large lattices $\xi_\infty$ may be replaced by  $L$.  
This substitution  then gives for the FSS of the 
susceptibility and the $j^{\rm{th}}$ Lee-Yang zeros,
\begin{eqnarray}
 \chi_L & \sim & L^{\frac{\gamma}{\nu}}\left({ \ln{L} }\right)^{\frac{ \nu \hat{\gamma}-\gamma \hat{\nu}}{\nu}} \,, \\
\label{FSSchi}
 r_j & \sim & L^{-\frac{\Delta}{\nu}}\left({ \ln{L} }\right)^{\frac{ \nu \hat{\Delta}+\Delta \hat{\nu} }{\nu}}
\label{FSSh1}
\,,
\end{eqnarray}
respectively.
The values for the correction exponents $\hat{\gamma}$ and $\hat{\nu}$ given in (\ref{leading}) and (\ref{SSLJscaling})
have been numerically established for the RBIM in Refs.~\cite{RoAd98,Aade96} 
and that for and $\hat{\nu}$ in the RSIM has been verified in Ref.~\cite{ShVa01}.
Therefore their confirmation in this setting as $\displaystyle{ \nu \hat{\gamma}-\gamma \hat{\nu} = 0}$ 
 serves as a useful check on the accuracy of our method at the logarithmic level.

\begin{table}
\caption{The estimates for the leading-exponent ratios $\gamma/\nu$ and $\Delta/\nu$ from fits to 
the scaling behaviour of the susceptibility and first Lee-Yang zeros as well as
estimates for $({\nu \hat{\gamma}-\gamma\hat{\nu}})/{\nu}$.
These estimates agree with the theoretical values, which are $7/4$, $15/8$ and $0$, respectively. 
}  
\begin {center}
\begin{tabular}{|l|r|r|r|}  \hline \hline
                                                                           & $p=0.88889 $ & $p=0.75$  & $p=0.66661$ \\
\hline
$\gamma/\nu$                                                     & $ 1.747  \pm 0.007$   &   $1.755 \pm 0.005$ &   $1.752 \pm 0.007 $  \\
$({\nu \hat{\gamma}-\gamma\hat{\nu}})/{\nu}$    & $-0.01   \pm 0.03$     &    $0.02 \pm 0.03 $ &   $0.01 \pm 0.03$     \\
$\Delta/\nu$                                                        & $ 1.879 \pm 0.004$    &   $ 1.878 \pm 0.006  $ &   $ 1.878 \pm 0.006 $  \\ 
\hline \hline
\end{tabular}
\end{center}
\label{tab3}
\end{table}
The scaling analysis begins with $p=0.88889$ (and $\beta=0.53781$).
A double logarithmic plot of $\chi_L $ against $L$ is presented in Fig.~1a, and a fit to all data points yields 
$\gamma / \nu = 1.747(7)$ in agreement with (\ref{leading}). 
The goodness of fit corresponds to  a $\chi^2$ per degree
of freedom ($\chi^2/{\rm{dof}}$) of $0.8$. 
The FSS correction exponent 
$\displaystyle{\left({ \nu \hat{\gamma}-\gamma \hat{\nu} }\right)/\nu}$ is extracted by a fit, corresponding to Fig.~1b,
of $\ln{\chi_L}-7/4 \ln{L}$ against $\ln{(\ln{L})}$, giving a slope $-0.01(3)$ (with $\chi^2/{\rm{dof}} \approx 0.8$) 
compatible with the expected value of zero from (\ref{leading}) and (\ref{SSLJscaling}).
These results are sumarized in Table~\ref{tab3} alongside the results of similar analyses at
$p=0.75$ ($\beta = 0.77125$) and $p=0.66661$ ($\beta = 1.10$). In each case, the $\chi^2/{\rm{dof}}$ indicates 
a good fit and compatibility with  the SSLJ theory  is firmly established.
\begin{figure}[t]
\vspace{7cm}
\includegraphics{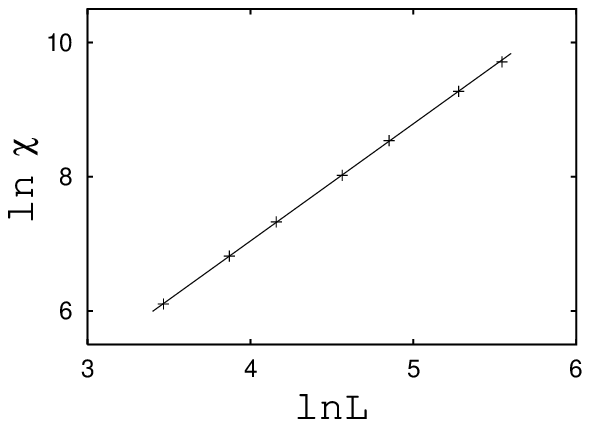}
\includegraphics{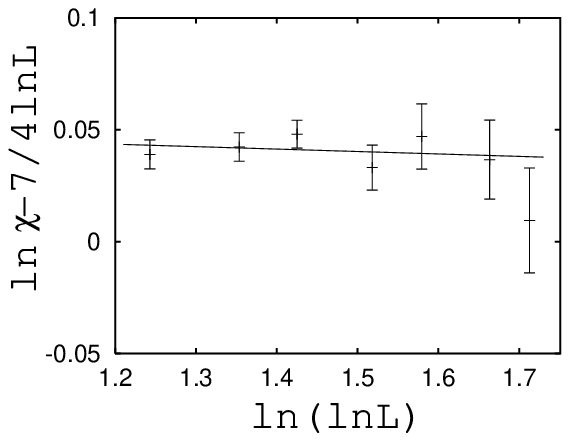}
\caption[a]{(a) FSS plot for $\chi_L$ at $p_c = 0.88889$. 
The slope gives an estimate for $\gamma / \nu$ of $1.747(7)$.
(b) Plot of  $\ln{\chi_L}-7/4 \ln{L}$ against $\ln{(\ln{L})}$ giving slope $-0.01(3)$, indicating no multiplicative logarithmic corrections to the 
FSS of the susceptibility.}
\vspace{-7.5cm} \hspace{1.3cm}
(a)
\hspace{-1.3cm} \vspace{7.5cm}
\vspace{-7.5cm} \hspace{8.5cm}
(b)
\hspace{-8.5cm} \vspace{7.5cm}
\label{fig:chi}
\end{figure}

Subleading corrections to  (\ref{FSSchi}) are expected to take the form \cite{HaTo08}
\begin{equation}
\chi_L = A  L^{\frac{\gamma}{\nu}}\left({ \ln{L} }\right)^{\frac{ \nu \hat{\gamma}-\gamma \hat{\nu}}{\nu}}
\left[{
1 + {\cal{O}}\left({
\frac{1}{\ln{L}}
}\right)
}\right]
 \,.
\label{step3}
\end{equation}
The amplitude $A$ of the leading term may firstly be estimated by fitting to $ \chi_L = A  L^{{\gamma}/{\nu}}$. 
A subsequent fit to the parameters governing corrections to scaling in (\ref{step3}) 
yields the estimate  $\displaystyle{\left({ \nu \hat{\gamma}-\gamma \hat{\nu} }\right)/\nu} = -0.03(2)$ (with $\chi^2/{\rm{dof}} \approx 0.8$)  for 
at $p=0.88889$.
I.e., the  inclusion of additive corrections does not lead to an improved estimate for the multiplicative logarithmic
exponents. 
(We have also tested additive corrections of the form $\displaystyle{\ln{\ln{L}}/\ln{L}}$ in place of $\displaystyle{1/\ln{L}}$
in (\ref{step3}). While this leads to small improvement over  (\ref{step3}), it also does not significantly affect the estimates for the 
exponents of the multiplicative logarithms.)
This observation holds for all quantities analysed below and for all values of $p$, and we henceforth
refrain from reporting on additive corrections.

The analysis of the leading FSS behaviour at $p=0.88889$  
of the first Lee-Yang zero of Ref.~\cite{RL} is reconfirmed (to higher precision)
in Fig.~2a, 
\begin{figure}[h]
\vspace{7cm}
\includegraphics{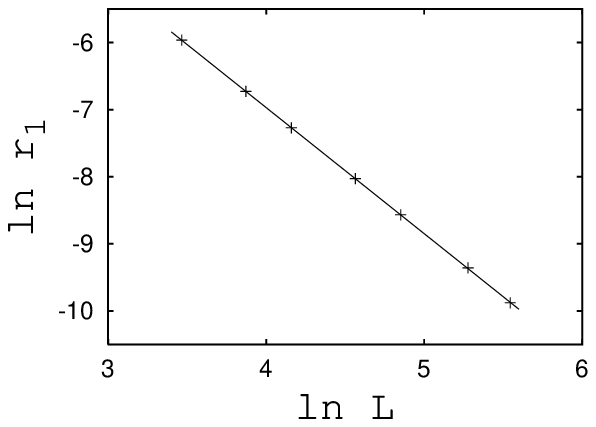}
\includegraphics{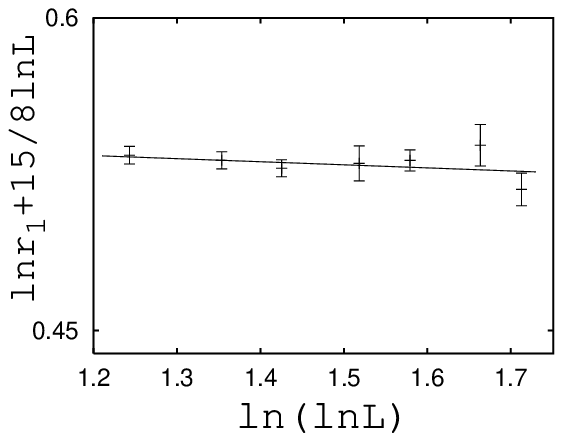}
\caption[a]{(a) FSS plot for the lowest Lee-Yang zero at $p_c = 0.88889$. 
The slope gives an estimate for $\Delta / \nu$ of $1.879(4)$.  
(b) Plot of  $\ln{r_1}+15/8 \ln{L}$ against $\ln{(\ln{L})}$ giving slope $-0.01(2)$, indicating the absence of 
multiplicative logarithmic corrections to the FSS of $r_1$.}
\vspace{-8.0cm} \hspace{1.3cm}
(a)
\hspace{-1.3cm} \vspace{7.5cm}
\vspace{-7.5cm} \hspace{8.5cm}
(b)
\hspace{-8.5cm} \vspace{7.5cm}
\label{fig:h1}
\end{figure}
where, ignoring corrections, the slope of the log-log plot yields $\displaystyle{\Delta/\nu = 1.879(4)}$,
compatible with the expected value of $15/8$.   The $\chi^2/{\rm{dof}}$ here is $0.7$.
While the corresponding fit at $p=0.66661$ yields $\displaystyle{\Delta/\nu = 1.878(6)}$,
 the equivalent fit (using all lattice sizes) for $p=0.75$ gives the estimate 
$1.883(4)$, which is two standard deviations from the theoretical value.
To avoid contamination due to scaling corrections, the smallest lattices may be removed from the analysis in standard fashion,
and compatibility with theory is indeed restored when the smallest pair  are removed from the fit;
the $L=64$ to $L=256$ data again yields $\displaystyle{\Delta/\nu = 1.878(6)}$
(with $\chi^2/{\rm{dof}}=1.8$).
These reconfirmed results are also summarized in Table~\ref{tab3}.

To test the  scaling relation (\ref{gaplog}) and the prediction (\ref{loggap}), 
$\ln{r_1}+15/8 \ln{L}$ is plotted against $\ln{(\ln{L})}$ in Fig.~2b for $p=0.88889$. A fit to all points yields slope
\begin{equation}
 \frac{ \nu \hat{\Delta}+\Delta \hat{\nu} }{\nu} = -0.01(2)
\,,
\label{slope}
\end{equation}
compatible with zero.
The $\chi^2/{\rm{dof}} $ here is $0.7$.  
With the by now established theoretical values for $\Delta$, $\nu$ and $\hat{\nu}$ from 
(\ref{leading}), (\ref{gap}) and (\ref{SSLJscaling}), this  yields
\begin{equation}
 \hat{\Delta} = - 0.95(2)
\,,
\label{952}
\end{equation}
a value compatible with (\ref{loggap}) and confirming (\ref{gaplog}) in the RSIM.
The analysis  at $p=0.66661$ 
gives  $\displaystyle{( \nu \hat{\Delta}+\Delta \hat{\nu} )/{\nu} = -0.01(3)}$,
comparable to (\ref{slope}) and compatible with theory
(the corresponding $\hat{\Delta}$ value is $-0.95(3)$).
At $p=0.75$ one finds the estimate $-0.04(2) $, using all lattice sizes. 
Again, compatibility  with theory is restored by dropping the two smallest $L$ values, 
where one again finds  $-0.01(3)$ ($\hat{\Delta} = -0.95(3)$).
These estimates for $\hat{\Delta}$ are summarized in Table~\ref{tab4}.

Accepting, now, that the values $\Delta = 15/8 = 1.875$ from (\ref{gap}) and
$ \hat{\Delta} = -{15}/{16} = - 0.9375 $ from (\ref{loggap}) are supported by the data for each 
dilution level, (\ref{a2}) and (\ref{uncommon}) give the scaling exponents for the specific heat. In particular, 
the logarithmic correction exponent is identical to that of the density of zeros,
\begin{equation}
 \hat{\alpha} = \hat{a}_2
\,.
\label{simple}
\end{equation}
Therefore $\alpha = 2 - 16a_2/15$ and $\hat{\alpha} = \hat{a}_2$ can be extracted from the scaling form   (\ref{G}) for the density of zeros, 
a form which is expected to hold for small enough $r$ (i.e., close to the critical region).

A robust method to determine the density of zeros from simulational data was developed in Ref.~\cite{density}.
Define the density of zeros along the singular line $r>r_{\rm{YL}}(t)$ as
\begin{equation}
 g_L(r) = L^{-d} \sum_j{\delta \left({ r-r_j(L) }\right)}
\,,
\end{equation}
where $r_j(L)$ is the position of the $j^{\rm{th}}$ zero  for a lattice of extent $L$. Here $j$ is called the index of the zero.
Integrating along the locus of zeros gives the cumulative density of zeros (or index density) to be
\begin{equation}
 G_L(r) = \int_0^r{g_L(s)ds}
 = \frac{j}{L^d} \quad {\mbox{for}} \quad r_j(L) < r < r_{j+1}(L)
\,,
\end{equation}
so that at a zero, it is given by the average
\begin{equation}
 G_L(r_j(L)) = \frac{2j-1}{2L^d}
\,.
\end{equation}
From (\ref{G}), this  may be fitted 
 to the form
\begin{equation}
 G(r) = a_1 r^{a_2} (\ln{r})^{\hat{a}_2}+ a_3
\,,
\label{fit}
\end{equation}
allowing for an additional parameter $a_3$
which determines the phase.
A value of $a_3$ greater or less than zero indicates that the system is in the broken or symmetric phase, respectively, so that
$a_3=0$ only at the transition point $t=0$.
The second criterion for a good fit is good data collapse.  
Note that this method does not allow an independent goodness-of-fit test \cite{density}.

\begin{figure}[t]
\vspace{7cm}
\includegraphics{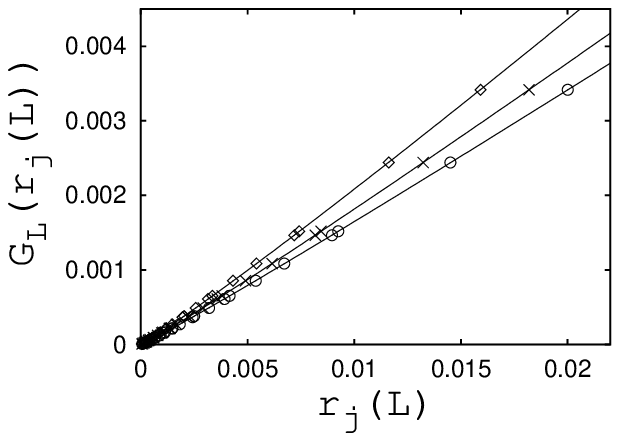}
\includegraphics{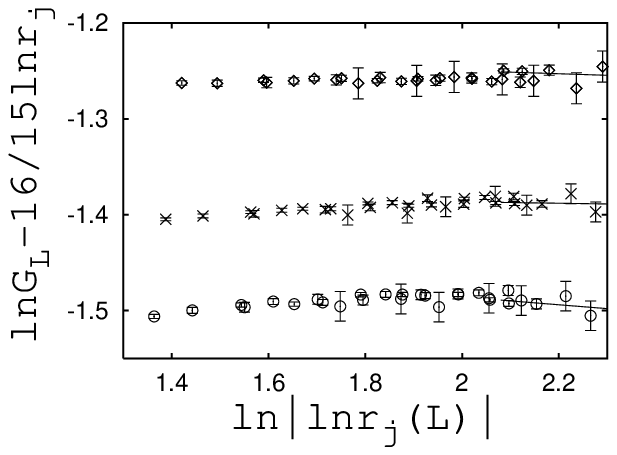}
\caption[a]{(a) The integrated density of zeros  at $p = 0.88889$ (`$\diamond$' symbols), $p = 0.75$ (`$\times$') and $p=0.66661$ (`$\circ$'). 
The excellent data collapse (four data points for each of seven lattice sizes are involved at each value of $p$)
and zero intercept indicate the correct transition point and the fitted curves
give $a_2$ compatible with the expected value $16/15$ ($\alpha = 0$).
(b) The absence of multiplicative logarithmic corrections in the integrated density of zeros indicates that 
the specific heat logarithm exponent $\hat{\alpha}=0$ independent of the degree of dilution. }
\vspace{-10.0cm} \hspace{1.5cm}
(a) \hspace{0.5cm}
\hspace{-2.0cm} \vspace{8.5cm}
\vspace{-7.5cm} \hspace{8.5cm}
(b)
\hspace{-8.5cm} \vspace{8.7cm}
\label{fig:G}
\end{figure}

The integrated density of zeros is plotted in Fig.~3a using the first four Lee-Yang zeros for lattices from size $L=32$ to $L=256$
($28$ points in all)
for each value of the dilution, demonstrating excellent data collapse in each case. 
For $p=0.88889$, three-parameter fits to (\ref{fit}), for small $r$, with $\hat{a}_2=0$ 
are supportive of the theoretical values of  $a_2=16/15$ and $a_3=0$; 
using the eight lowest data points, one obtains  $a_3 = 0.000\,000\,2(4)$ and $a_2 =1.076(16)$.
The latter results corresponds to the estimate $\alpha = -0.02(3)$, from (\ref{a2}). 
The corresponding results in the $p = 0.75$ and $p=0.66661$ cases are 
$a_3=-0.000\,000\,2(3)$, $a_2=1.062(10)$ ($\alpha = 0.01(2)$) and
$a_3=-0.000\,000\,1(4)$, $a_2=1.066(15)$ ($\alpha = 0.00(3)$), respectively.
These results for $\alpha$ are gathered in Table~\ref{tab4}.

We now accept that $a_3$ is indeed zero and the theoretical value $a_2=16/15$ ($\alpha = 0$) from (\ref{a2}) holds  for each dilution.
Potential multiplicative logarithmic corrections are detected by plotting 
$\ln G - 16/15 \ln r$ against $\ln{(\ln{r})}$ in Fig.3b. 
A fit to all data points 
for $p=0.88889$ gives $\hat{\alpha} = \hat{a}_2 =  0.012(3)$, which is four standard deviations from the
theoretical value of zero. However, focusing on the scaling region closer to the origin establishes
compatibility with the theory. For example, fitting to the lowest eight data points yields 
$\hat{\alpha} = \hat{a}_2 = -0.02 (5)$. 
The equivalent results for $p=0.75$ and $p=0.66661$ are 
$\hat{\alpha} = \hat{a}_2 = -0.01(3)$ and $-0.04(5)$, respectively. 
The corresponding fits are depicted in Fig.~3b and the estimates for $\hat{\alpha}$ are summarized in Table~\ref{tab4}.
These values constitute  numerical evidence that $\alpha= \hat{\alpha}=0$,
independent of dilution and in favour of strong universality.
\begin{table}
\caption{The estimates for the specific heat exponent $\alpha$ and logarithmic-correction exponents $\hat{\Delta}$ and $\hat{\alpha}$. 
These are in agreement with the theoretical values from the strong universality hypothesis, namely 
$\alpha=0$, $\hat{\Delta}=-15/16$ and $\hat{\alpha}=0$. 
}  
\begin {center}
\begin{tabular}{|l|r|r|r|}  \hline \hline
                                                                           & $p=0.88889 $ & $p=0.75$  & $p=0.66661$ \\
\hline
$\hat{\Delta}$                                                     & $ -0.95  \pm 0.02$   &   $-0.95 \pm 0.03$ &   $-0.95 \pm 0.03 $  \\
$\alpha$                                                     & $ -0.02  \pm 0.03$   &   $0.01 \pm 0.02$ &   $0.00 \pm 0.03 $  \\
$\hat{\alpha}$                                                     & $ -0.02 \pm 0.05$    &   $-0.01 \pm 0.03$ &   $-0.04 \pm 0.05 $  \\ 
\hline \hline
\end{tabular}
\end{center}
\label{tab4}
\end{table}

\section{Conclusions}
\label{Simulation}
\setcounter{equation}{0}

The debate regarding the critical behaviour of the disordered  Ising model in two dimensions has persisted for over
thirty years (most recently in \cite{HaTo08,HaMa08}). 
Here the SSLJ prediction \cite{Boris,SL,JuSh96} for the multiplicative  logarithmic corrections for the scaling behaviour of the 
susceptibility has been reconfirmed through a careful FSS analysis. 
In addition to this, the  Lee-Yang zeros have been determined to high accuracy
and their logarithmic corrections verified for the first time.

The scaling behaviour of the specific heat in the site-diluted version of the model has been particularly 
difficult to pin down directly, and fits to the measured specific heat as a function of the double 
logarithm of the lattice extent \cite{AnDo90WaSe90TaSh94,HaTo08,BaFe97,SeSh98} have been claimed not
to be unambiguous  \cite{Ki00b}. Here an alternative approach has been taken, involving the
density of  Lee-Yang zeros. Using scaling relations \cite{KeJo06a,KeJo06b} to connect to the even sector of 
the model, the specific-heat scaling and correction exponents are clearly determined.
Since the simulations are performed at three different values of the dilution (some quite large)
the analyses presented herein for the susceptibility, the individual Lee-Yang zeros and for their densities,
are unambiguously supportive of the  strong scaling hypothesis.

~\\~\\
\noindent
{\bf{Acknowledgements:}}
We thank Boris Shalaev for e-mail correspondences.
This work has been partially supported by MEC through contracts No.
FIS2006-08533 and FIS2007-60977.

\bigskip
%

\end{document}